\documentclass{elsarticle}
\usepackage{xspace}

\usepackage{siunitx}
\usepackage[hidelinks]{hyperref}
\usepackage[color=yellow]{todonotes}
\usepackage[capitalise]{cleveref}

\newcommand{\xcore}{xCORE\xspace}
\newcommand{\level}{level\xspace}

\newcommand{\levels}{levels\xspace}
\newcommand{\hcir}{HC IR\xspace}
\newcommand{\llvmir}{LLVM IR\xspace}

\newcommand{\isa}{ISA\xspace}
\newcommand{\ciao}{Ciao\xspace}
\newcommand{\ciaopp}{CiaoPP\xspace}

\newcommand{\tclk}{T_{\text{clk}}}
\newcommand{\pbase}{P_{\text{b}}}

\journal{MICPRO}

\begin{document}
\begin{frontmatter}
\title{ENTRA: Whole-Systems Energy Transparency}

\author[unibris]{Kerstin~Eder}
\author[roskilde,imdea]{John~P.~Gallagher\corref{mycorrespondingauthor}}
\cortext[mycorrespondingauthor]{Corresponding author}
\ead{jpg@ruc.dk}
\author[imdea,csic]{Pedro~L\'{o}pez-Garc\'{i}a}
\author[xmos]{Henk~Muller}
\author[imdea]{Zorana~Bankovi\'{c}}
\author[unibris]{Kyriakos~Georgiou}
\author[imdea]{R\'{e}my~Haemmerl\'{e}}
\author[imdea,upm]{Manuel~V.~Hermenegildo}
\author[roskilde]{Bishoksan~Kafle}
\author[unibris]{Steve~Kerrison}
\author[roskilde]{Maja~Kirkeby}
\author[imdea]{Maximiliano~Klemen}
\author[roskilde]{Xueliang~Li}
\author[imdea]{Umer~Liqat}
\author[unibris]{Jeremy~Morse}
\author[roskilde]{Morten~Rhiger} 
\author[roskilde]{Mads~Rosendahl}

\address[unibris]{University of Bristol, UK}
\address[roskilde]{Roskilde University, Denmark}
\address[imdea]{IMDEA Software Institute, Spain}
\address[xmos]{XMOS Ltd., Bristol, UK}
\address[csic]{Spanish Council for Scientific Research, Spain}
\address[upm]{Technical University of Madrid, Spain}

\begin{abstract}

 Promoting energy efficiency
to a first class system design goal is an important research
challenge.  Although more energy-efficient hardware can be
designed,  it is software that controls the hardware; for a given system
the potential for energy savings
is likely to be much greater at the higher levels of abstraction in the system
stack. Thus the greatest savings are expected from energy-aware
software development, which is the vision of the EU ENTRA project.
This article presents the concept of
\emph{energy transparency} as a foundation for
energy-aware software development. 
We show how energy modelling of hardware is combined with static analysis
to allow the programmer to understand the energy consumption of a program without
executing it, thus enabling exploration of the design space taking energy into consideration.
The paper concludes by  summarising the
current and future challenges identified in the ENTRA project.

\end{abstract}
\end{frontmatter}

\section{Introduction}
Energy efficiency is a major
concern in systems engineering. 
The
EU's Future and Emerging Technologies MINECC programme aims to ``lay the foundations for radically
new technologies for computation that strive for the theoretical
limits in energy consumption.'' The research objectives range 
from physics to software; they include, among others, new elementary devices, 
as well as
``software models and programming methods supporting the strive
for the energetic limit.''
The ENTRA project, {\small\tt\url{entraproject.eu}}, addresses the
latter objective; we focus on energy transparency, which
we regard as a key prerequisite for new
energy-aware system development methods and tools.

\begin{figure}
 \begin{center} \includegraphics[width=0.8\columnwidth]{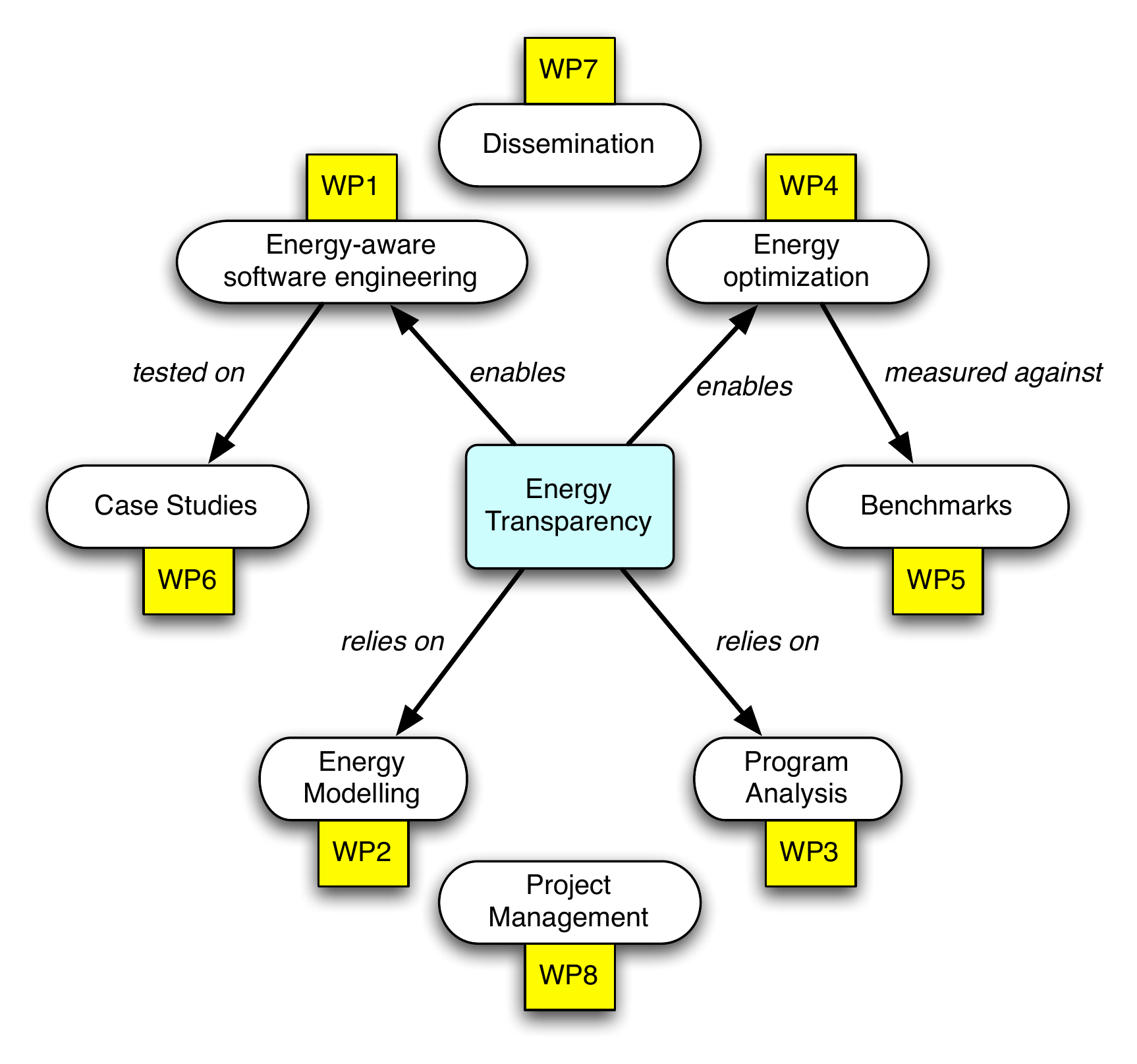}\end{center}
  \caption{Overview of the ENTRA project work plan}
  \label{fig:entra-wp}
\end{figure}

The ENTRA project ran from October 2012 to December 2015 (39 months) and was funded
by the European Commission under the 7th Framework Programme.  The consortium contained
three research institutions and one industrial partner specialising in the design of
advanced multicore microcontrollers (XMOS \xcore).
The overview of the project structure is shown in Figure \ref{fig:entra-wp}.  The foundations for
the central concept of energy transparency were developed in two work packages (WP2 and WP3) on
energy modelling and energy analysis respectively. Energy transparency enables 
energy optimisations, studied in WP4.  WP1 concerned the development of tools and techniques applicable in energy-aware software
development.  Finally there were work packages dealing with benchmarking, evaluation, dissemination and project management.
This paper summarises mainly the outcomes of work packages WP1, WP2, WP3, WP4 and WP6.  The public deliverables of the
project are all available on the project website \texttt{http://entraproject.eu}.

After this introduction, we discuss the two main areas of research
supporting energy transparency. \cref{energy-models}
presents approaches for building models of software energy
consumption at different levels of abstraction.
\cref{energy-analysis} contains an overview of static resource
analysis techniques, showing how an energy model can be used in
analysis of a program's energy consumption.
\cref{conclusion} summarises the role of energy transparency
in energy-aware software development, discusses the achievements 
in the project so far, and outlines current challenges and directions for future research.

\subsection{Energy-Aware Computing}\label{sec1}
Energy-aware computing is a research challenge that requires
investigating the entire system stack from application software and
algorithms, via programming languages, compilers, instruction sets and
micro architectures, to the design and manufacture of the hardware.
This is because energy is consumed by the hardware performing
computations, but the control over the computation ultimately lies
within 
the applications running on
the hardware. 
While hardware can be
designed to save a modest amount of energy, the potential for savings
is far greater at the higher levels of abstraction in the system
stack. 
An estimate from Intel \cite{Edwards:lssmlps} is that
energy-efficient software can realize savings of a factor of three to five beyond what can be achieved through energy efficient hardware.
Roy and Johnson
\cite{Roy_Johnson_1997} 
list five objectives that ``help make software design
decisions consistent with the objectives of power minimization'': match the algorithm to the hardware; minimize 
memory size and expensive memory
accesses; optimise the performance, making
maximum use of available parallelism; take advantage of hardware
support for power management; and select instructions, sequence them,
and order operations in a way that minimizes switching in the CPU and
datapath.  To achieve these objectives requires the programmer and/or the tools
to understand the relationship between code and
energy usage. 
Energy Transparency aims to enable exactly this.
\subsection{Energy Transparency}
The concept of energy transparency is
at odds with the
trend in modern software engineering - the desire to abstract away
machine-level details using high-level languages, abstract data types
and classes,  libraries and layers of
interpretation or compilation, in the interests of portability, 
programmer productivity, understandability and software reuse. By
contrast, energy transparency requires making visible how software impacts
on energy consumption when executed on hardware. Availability of this
information enables system
designers to find the optimal trade-off between performance, accuracy,
and energy usage of a computation. 
To achieve energy transparency, models of how energy is consumed
during a computation are required.  As will be discussed in
\cref{energy-models}, such models can
be established at different levels of abstraction, ranging from models
that characterize individual functional hardware blocks
\cite{wattch:2000}, via Instruction Set Architecture (ISA)
characterization models
\cite{TiwariWolfeInstructionLevelPowerAnalysi:1996,Steinke2001,DBLP:journals/tecs/KerrisonE15}, to
models based on 
intermediate representations used by the compiler 
\cite{Brandolese2011,Georgiou2015arXiv}.
The final energy models 
provide information that feeds into static resource usage analysis
algorithms 
\cite{DBLP:journals/cacm/Wegbreit75,Rosendahl89,granularity,low-bounds-ilps97,resource-iclp07,NMHLFM08,plai-resources-iclp14,AlbertAGP11a}, 
where they represent the
energy usage of elementary parts of the computation. This is discussed in \cref{energy-analysis}.

\section{Energy Modelling}\label{energy-models}

Energy models
can rely on information at several possible abstraction levels, from gate-level
hardware description, through functional block and Instruction Set Architecture
(ISA), up to performance counter or transaction based abstractions. Energy models at higher
levels tend to be faster to use, but have lower accuracy than models at lower levels of abstraction. In ENTRA,
the aim was to provide accurate modelling that can be exploited through analysis
that is applied in order to estimate the energy consumed by software.

\subsection{Defining and constructing an energy model}\label{ss:xcore-e-model}

The ISA is a practical level of abstraction for energy modelling, because it expresses
underlying hardware operations and their relationship with the intent of the
software. Constructing a model at this level gives us the following benefits:
 energy costs can be attributed directly to individual machine instructions as output by the back end of the compiler;  
 instruction properties and energy consumption are strongly correlated, e.g.\ energy consumption typically increases with increasing numbers of operands;
and machine instructions can be traced back to the original source code
    statements written by the software developer, as well as to various
    intermediate representations.

However,
energy modelling at the ISA level requires additional effort in order to produce
useful models:
 instruction costs must be captured through a profiling suite and
    measurement of device power;
    in addition, indirect or statistical approaches are required to
    characterise instructions that cannot be profiled through direct measurements.
Furthermore, for multi-threaded architectures other properties such as the cost of running multiple threads and the cost of idle periods must be determined.

Our target architecture for energy modelling and analysis is the XMOS \xcore embedded microcontroller~\cite{XS1-Architecture}.
Beyond offering timing-deterministic instruction execution, the \xcore is
hardware multi-threaded and comes in a variety of multi-core configurations.
The \xcore architecture is simple by design and, thus, ideal to investigate the
advanced energy modelling and static analysis techniques required to achieve
energy transparency. 
The techniques we developed are readily transferable to other deeply embedded,
cache-less, IoT-type processors such as those in the ARM Cortex M series or the Atmel AVR. The
fact that the \xcore offers multi-threading made it a particularly interesting target for the ENTRA project.

We have shown that in the \xcore the number of active threads has an impact
upon energy consumption~\cite{DBLP:journals/tecs/KerrisonE15}. As such, the model must take this into account.
Traditional ISA-level models, such as that of~\cite{Tiwari1994a}, can attribute
energy costs simply to instructions, the transitions between instructions, and
any additional effects that impact on energy consumption, such as cache hits and misses. Although we build on this
principle, parallelism has to be considered, yielding a more complex model
equation for $E_{\text{p}}$, the energy consumed by a program $\text{p}$:

\vspace*{-2mm}
\begin{equation}
E_{\text{p}} = \pbase N_{\text{idl}}\tclk + \sum_{t=1}^{N_{t}}{\sum_{i \in \text{ISA}}{\left(\left(M_{t}P_{i}O + \pbase\right) N_{i,t} \tclk\right) }}
\label{eq:coremodel}
\end{equation}

In~\cref{eq:coremodel} the energy consumption is split into two parts,
capturing idle and active processor behaviour, respectively. For the former, we
consider the \emph{base processor idle power}, $\pbase$, that is present even
when the device is waiting on external events, multiplied by the the number of
cycles with no active threads, $N_{\text{idl}}$, and the clock period, $\tclk$.
For the latter, individual instruction costs are accounted for based on their
costs $P_i$, as well as an aggregated inter-instruction overhead, $O$, and a
parallelism scaling factor, $M_t$, determined by the number of active threads $t$. This is
calculated for each ISA instruction $i$, and multiplied by the number of
occurrences in the target program at that particular level of parallelism,
$N_{i,t}$, as well as the clock period, $\tclk$.

Model parameters are separated into two groups. Values for the first group of constants are
obtained by profiling the processor for a fixed clock period $\tclk$, yielding
the base power $\pbase$, inter-instruction overheads $O$, per-instruction costs
$P_i$ and parallelism scaling $M_{t}$; all measured in $mW$. The second group
must be determined through analysis of the target program. These include the
number of idle cycles, $N_{\text{idl}}$, the number of threads, $N_{t}$, in the
program, and the instruction counts, $N_{i,t}$, for each instruction $i$ and
number of active threads $t$. 
If the analysis can produce these values, \cref{eq:coremodel} can be used to
estimate program energy. We have demonstrated various simulation- and static
analysis-based methods that follow this principle.

\begin{figure}
 \begin{center} \includegraphics[width=1.0\columnwidth, clip, trim=2.3cm 0.9cm 2.6cm 1.3cm]{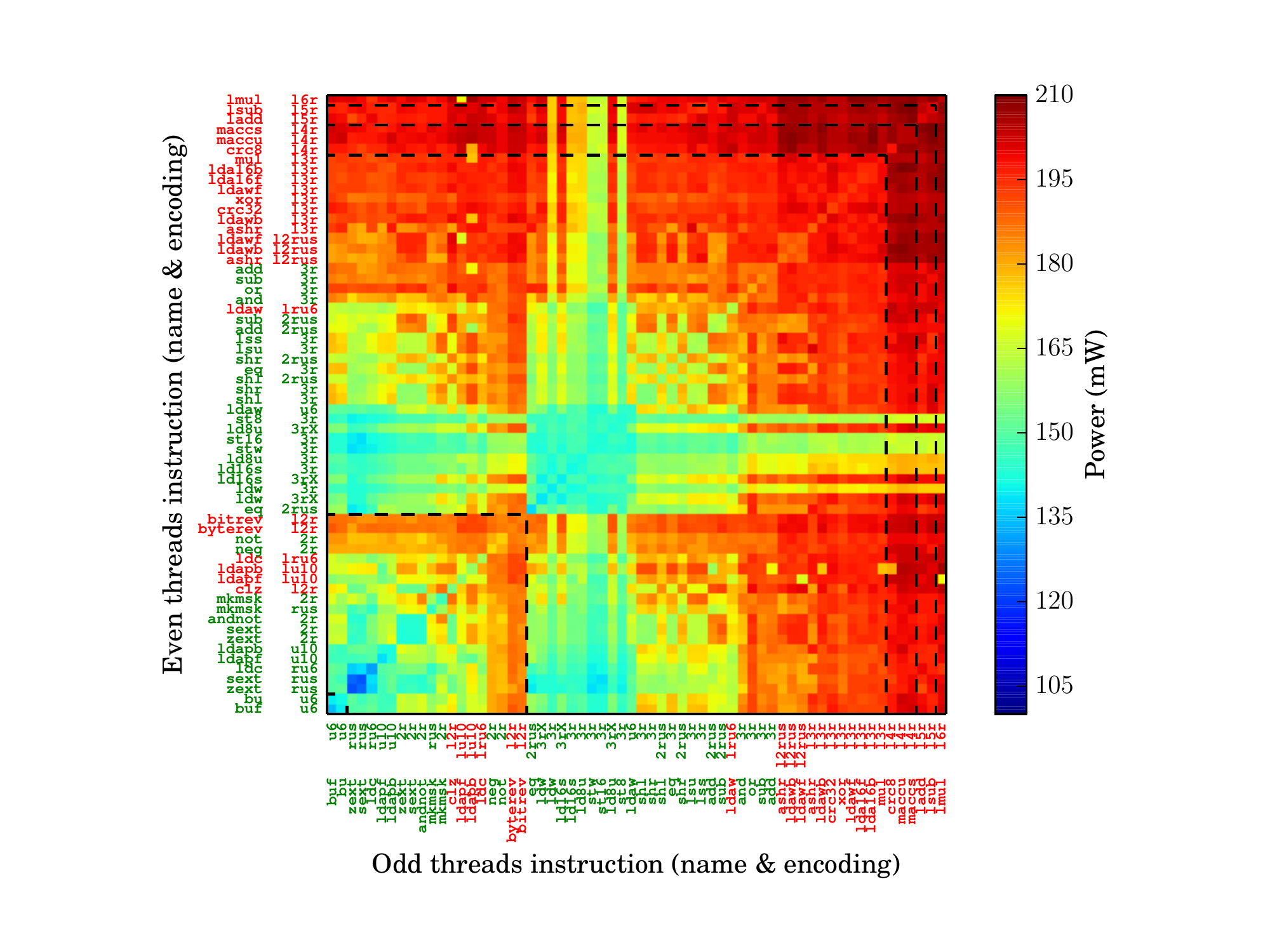}\end{center}
   \caption{Power dissipation of multi-threaded instruction interleaving in the XMOS \xcore processor. Dashed lines denote a change in operand count, axis label colour indicates 16-bit (green) and 32-bit (red) instruction encoding.}
  \label{fig:heatmap}
\end{figure}

To illustrate instruction profiling, an example \emph{heat map} representing
the device power from interleaving a selected subset of data manipulation
instructions is shown in \cref{fig:heatmap}. The profiling framework executes
tightly coupled threads through the \xcore pipeline, with random, valid operand
values to produce an average power estimate for each instruction. Random input
data is shown to cause higher power dissipation than more constrained data that
would be found in real-world programs~\cite{pallister2015data}, e.g.\ due to
data dependencies, thus creating a modest over-estimate in most cases.

Instruction profiling can only be used to determine the costs of instructions that
can be executed repeatedly and in succession. The cost of other instructions can be
estimated with a generic average or grouped by operand
count~\cite{DBLP:journals/tecs/KerrisonE15}, or more accurately through a
regression tree approach that identifies the most significant of a set of
features, including instruction length, whether memory is accessed and others, to
find the most similar directly profiled instruction~\cite{2015arXiv150902830K}.
The latter is the most accurate of the three approaches and adds no significant
modelling overhead.

This form of model can be used by various analysis methods. An Instruction Set
Simulation (ISS) can produce an instruction trace from which the instruction
counts and thread activity can be determined. A cycle-accurate ISS will achieve
the best model accuracy. Alternatively, per-thread instruction execution
statistics can be used instead to extrapolate the model terms. This is faster
than producing a full trace, but can increase estimation error. It has been
shown to yield an acceptable $\pm$\SI{10}{\percent} margin in
benchmarks~\cite{DBLP:journals/tecs/KerrisonE15}. 

As we will see in Section \ref{energy-analysis}, the program-dependent terms 
in the model can
be calculated by static analysis, removing the simulation step and allowing
rapid design-space exploration, as well as parameterised, bounded energy
estimations.  The values of $N_{i,t}$ and $N_{\text{idl}}$ can be analysed as 
functions of the input state.  This allows the derivation of energy functions characterising
the energy consumption of all possible runs of a program, rather than a specific run or set of runs.

In our experiments, tracing was up
to two orders of magnitude slower than statistics-only simulation, the latter
typically taking less than one minute. However, this can be mitigated by
analysing single functions or blocks, ignoring other parts of the trace, and
terminating as soon as the blocks of interest have been executed, making trace
simulations take less than a minute in most of the cases we observed.
Static analysis can, of course, achieve faster results, by analysing the same
code blocks without simulation.

A key prerequisite for achieving high accuracy in energy modelling at the ISA
level is a predictable, time-deterministic architecture, where the instruction
set gives an accurate view of how the processor will behave. 
The \xcore's thread scheduling and cache-less SRAM memory subsystem, together
with the absence of performance enhancing complexity in the micro-architecture,
enabled us to achieve a very accurate model, which is essential to obtaining accurate energy predictions. 
Thus, as for worst case execution time (WCET) analysis, the results of energy
modelling and energy consumption analysis are influenced by processor
architecture, and predictability determines the accuracy achievable as well as
the complexity of the modelling and analysis
techniques~\cite{influence-wilhelm-2003}. 

\subsection{Multi-core energy modelling}

For a multi-core system the ISS must accurately simulate the network
behaviour in order to capture the timing and link-traversal of data. This
allows accurate estimation of communicating multi-core
processes~\cite{2015arXiv150902830K}. Static analysis must provide similar
characterisation, therefore instruction execution, network behaviour and the
flow of communication between processes, must be predictable in order to enable
energy transparency.
The single-core, multi-threaded processor model achieves an
average error of \SI{2.7}{\percent} over a suite of single- and multi-threaded
benchmarks, with a standard deviation of \SI{4.4}{\percent}. The multi-core
model demonstrates an average energy estimation error of \SI{-4.9}{\percent}
with a standard deviation of \SI{3.9}{\percent}. Models with less than \SI{10}{\percent} error
provide suitable accuracy for energy estimations.

\subsection{Energy modelling at higher levels of software abstraction}
\label{subsec:mapping}

Performing static analysis at the ISA level can benefit from good accuracy
due to its closeness to the hardware, but it can suffer from a loss of
useful information such as program structure and
types~\cite{isa-energy-lopstr13-final}. A good compromise is found by
modelling at the Intermediate Representation (IR)
used by compilers, 
where program information is preserved. Since the compiler is the natural
place for optimisations, modelling and predicting the energy consumption at
IR level could therefore enable energy specific optimisations.

Using a mapping technique, we lifted 
the energy model in~\cref{eq:coremodel} to the IR level of the
compiler~\cite{Georgiou2015arXiv},
implemented within the LLVM tool
chain~\cite{LattnerLLVM2004}.  
 ISA level energy
models can thus be propagated up to LLVM IR level, allowing energy
consumption estimation of programs at that level. This enables
static analysis to be performed at a higher level than ISA, thus
making energy consumption information accessible directly to the compiler
and optimiser.

The mapping technique determines the energy characteristics of LLVM IR
instructions. It provides on-the-fly energy characterization that takes into
consideration the compiler behavior, control flow graph (CFG) structure, types and other aspects
of instructions. 
Taking into account such information
improved the accuracy of the LLVM IR characterization significantly. The
experimental evaluation demonstrated that the mapping technique allowed for
energy consumption transparency at the LLVM IR level, with accuracy keeping
within 1\% of ISA-level estimations in most cases~\cite{Georgiou2015arXiv}.

In principle, the same mapping technique may be used to map the energy consumption
of programs to even higher levels, such as the source code. However, a fine grained
characterization for each line of source code using this method is impractical due
to the numerous transformations and optimisations introduced by the compiler and the loss of
accuracy resulting from the difficulty of associating energy consumption costs obtained at lower levels to source code lines.

An alternative approach to building a source-level energy model was
investigated in~\cite{2015arXiv151004165L}.  The target language here was Java
on the Android platform; any attempt to map a lower-level energy model up to
the source code would need to deal explicitly with the Java virtual machine as
well as operating system layers, a highly impractical strategy.  Instead, the
basic  energy-consuming operations from  the  source code are identified and
the correlations to energy  costs are found by measuring energy consumption in
a large number of execution cases and analyzing the results using techniques
based on regression analysis.  The resulting energy model of the basic
operations implicitly includes the effect of all the layers of the software
stack down to the hardware. The approach is inherently approximate;
nevertheless such an approach may be the only feasible one in complex software
stacks, when source-level energy models are needed, for instance to give the
source-code programmer an energy profile of the code under development. 

\section{Static Analysis of Energy Consumption}\label{energy-analysis}

Static analysis is the other key component of energy transparency.
It infers
information about energy consumed by programs 
without
actually running them.
As with energy models, analysis can be performed on
program representations 
at different levels 
in the
software stack, ranging from source code (in different programming
languages) to intermediate compiler representations (such as
\llvmir~\cite{LattnerLLVM2004}) or ISA.

Static analysis at a given \level consists of reasoning about
the execution traces of a program 
at that \level, in order to infer 
information (among other things) about 
how many times certain basic elements
of the program 
will be executed.
The role of the energy model 
is to provide information
about the
energy consumption of 
such basic elements;
it is used by the analysis
to infer information about energy consumption of 
higher-level
entities such as procedures, functions, loops, blocks and the whole program.

Analysis can also be performed 
at a given software level using energy models for 
a lower
\level. Such a
model needs to be mapped upwards 
to the higher level, as described in \cref{subsec:mapping}.
The information inferred by static analysis at a lower \level can also be
reflected upwards to a higher level using suitable mapping
information.

In the ENTRA project, this approach has been applied to the static
analysis 
of XC programs running on xCORE architectures.  However, our
framework is language- and architecture-neutral.  
We will return to this in \cref{conclusion}.

\subsection{Analysis/modelling trade-off}
\label{sec:analysis-model-trade-off}

\begin{figure}[ht]
\centering
\centerline{\includegraphics[scale=0.45]{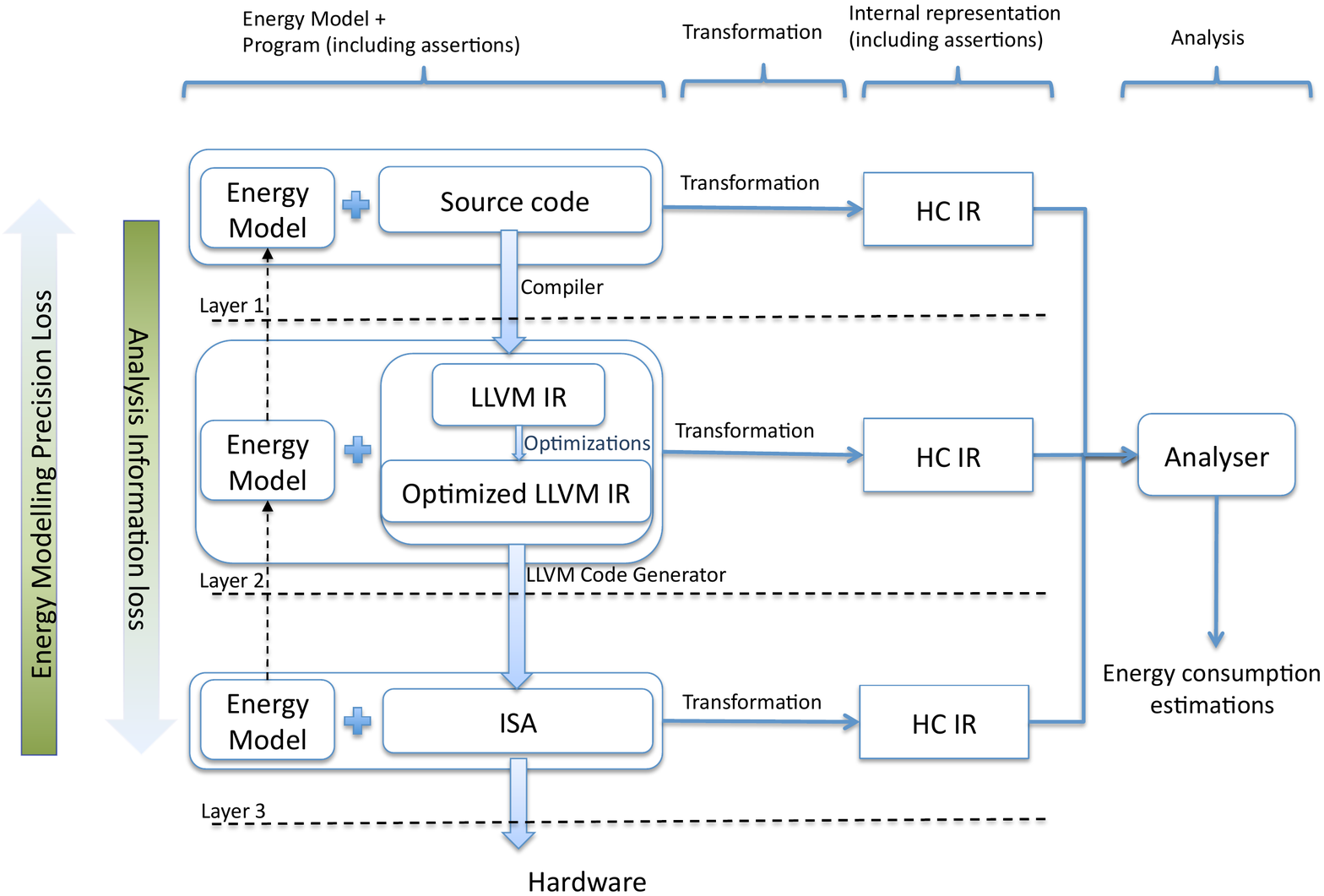}}
\vspace{-0.7cm}
\caption{Analysis/modelling \level trade-off and potential choices \cite{isa-vs-llvm-fopara}}
\label{analysis-model-choices-overview}
\end{figure}

Our hypothesis was that
the choice of \level affects the accuracy of the energy models and the
precision of the analysis in opposite ways: energy models at lower
\levels 
will be more precise than at higher
\levels 
while
at lower \levels more program structure and data 
structure
information is lost, 
which often implies
a corresponding loss of accuracy in the analysis.
This hypothesis about the analysis/modelling \level trade-off (and
potential choices) is illustrated in
Figure~\ref{analysis-model-choices-overview}.

In ENTRA we have explored different points in this space
of combinations of analysis and modelling.
Our
experimental results~\cite{isa-vs-llvm-fopara} confirm that the
expected trade-off exists, but also suggest that performing the static
analysis at the \llvmir \level is a good compromise.
\llvmir is close enough to the source code \level to preserve most of
the program information needed by the static analysis, whilst 
close enough to the ISA \level to allow the propagation of
the ISA energy model up to the \llvmir \level without significant loss
of accuracy. 

\subsection{Information inferred by analysis}
\label{sec:kind-of-info}

The information inferred by the analyzers is guided by its final use:
program optimisation, verification, helping energy-aware software
developers to make design decisions, and so on. For example, they can
infer safe approximations, namely upper and lower bounds, on the energy
consumed by the program or parts of it. These approximations can be
functions parametrised by the sizes of the input data and other hardware
features such as clock frequency and voltage. The analyzers can then
infer concrete values of the parameters that yield the worst-case energy
consumption of the program or its parts.

Static energy profiling~\cite{staticprofiling-flops} determines the
distribution of energy usage over the parts of the code. This can be
very useful to the developer, showing which parts of the program are the
most energy-critical. Some functions or blocks in the program are
perhaps not particularly expensive in energy in themselves but are
called many times. Such parts are natural targets for optimisation,
since there a small improvement can yield considerable savings. 

Note that the safety of bounds depends on energy models giving safe
bounds for each instruction.  This is a challenging problem which is
discussed in~\cref{conclusion}.  Safe bounds are vital for applying
energy analysis to verifying or certifying energy consumption.

\subsection{A generic resource analysis framework}

\label{sec:generality}

The resource analysis framework that we have developed
is parametric with respect to \emph{resources} and \emph{programming
  languages}.  Regarding resources, the common assertion
language
allows the definition of different resources and how basic
components of a program affect the use of such resources. More
concretely, it allows the encoding of different energy models for specific
hardware architectures, and, in particular, the energy models
developed for the xCORE architecture at the \llvmir and ISA levels,
described in~\cref{energy-models}.

Regarding  programming languages,
we differentiate between the \emph{input language} (which in ENTRA
 can currently be XC source, \llvmir, or ISA) and the
\emph{intermediate semantic program representation}, which is what the
resource analysis actually operates on. We use \emph{Horn Clauses} as
the intermediate program
representation,  referred to  as the ``\hcir.''  
A transformation is performed from each supported \emph{input language}
into \hcir, which is then passed to the resource analysis.
We have explored different approaches for this transformation.
One approach is to perform a \emph{direct} transformation into \hcir,
and this has been applied both to ISA and \llvmir
code~\cite{isa-energy-lopstr13-final,isa-vs-llvm-fopara}. Another
approach consists in producing the \hcir by applying partial
evaluation techniques to instrumented interpreters that directly
implement the semantics of the language to be
analysed~\cite{HenriksenG06}.  In both cases, the resulting \hcir
programs are analysed by the \ciaopp tool
(see~\cref{sec:horn-clause-analysis}).

Horn Clauses 
offer 
several
features that make
them convenient for analysis~\cite{decomp-oo-prolog-lopstr07}. For
instance, this representation inherently supports Static Single
Assignment (SSA) and recursive forms.  There is a current
trend favouring the use of Horn Clause programs as intermediate
representations in analysis and verification tools~\cite{DBLP:conf/tacas/GrebenshchikovGLPR12,DBLP:conf/fm/HojjatKGIKR12,z3,hcvs14}.

Using the generic \hcir representation, the assertion language 
and the \ciaopp analysis tools,
we have instantiated
the general framework to produce a series of 
concrete energy analyzers which have allowed us to study 
the advantages and limitations of different techniques as well as the
trade-offs implied by different choices of analysis and energy
modelling levels.

\subsection{A common assertion language}
\label{sec:common-assertion-language}

We have defined a common assertion language as a vehicle for
propagating energy-related information throughout the system \levels,
and for communication among the different analysis, verification, and
optimisation tools, and the actors involved in
software development. This assertion language allows
expressing energy models for different architectures, writing energy
consumption specifications, describing energy consumption of
components that are not available at analysis time, expressing
analysis results, and 
ensuring interoperability.
We refer the reader to~\cite{hermenegildo11:ciao-design-tplp} and its
references for a full description of the \ciao assertion language that
is the basis for the assertions used in the \hcir, and the (internal)
common assertion language of ENTRA~\cite{entra-d2.1}.  The ENTRA
common assertion language also includes a front end to express energy
specifications and other energy related information in the XC source
code.

\subsection{Energy analysis using \ciaopp}

\label{sec:horn-clause-analysis}

The input to the \ciaopp parametric static resource usage
analyzer~\cite{resource-iclp07,NMHLFM08,plai-resources-iclp14} is the
\hcir, along with assertions in the common assertion language
 expressing the
energy model for \llvmir blocks and/or individual \isa instructions,
and possibly some additional (trusted) information.
The analyzer is based on
an approach in which recursive equations (cost relations),
representing the cost of running the program, are extracted from the 
program and solved, obtaining upper- and lower-bound cost functions 
(which may be polynomial, exponential or logarithmic)
in terms of the program's
inputs~\cite{DBLP:journals/cacm/Wegbreit75,Rosendahl89,granularity,low-bounds-ilps97,AlbertAGP11a}. 
  
These output cost functions 
express the energy consumption for each block in the \hcir,
which can be mapped directly back to the language represented by the \hcir.

The generic resource analysis engine is fully based on \emph{abstract
  interpretation}~\cite{plai-resources-iclp14},
and defines the resource analysis itself as an \emph{abstract domain}
that is integrated into the PLAI abstract interpretation
framework~\cite{ai-jlp} of \ciaopp. This brings in features such as
\emph{multivariance}, efficient fixpoints, and assertion-based
verification and user interaction. The setting up and solving
of recurrence relations for inferring closed-form functions
representing bounds on the sizes of output arguments and the resource
usage of the predicates in the program are integrated into the PLAI
framework as an abstract operation.

\subsection{Direct energy analysis of LLVM IR}
\label{sec:llvm-static-direct}

As mentioned, \llvmir offers a good trade-off between analyzability and
accuracy. 
In addition to
using a generic approach based on \ciaopp and a translation to
\hcir, the ENTRA project has 
experimented with an approach that uses
similar analysis techniques but operates directly on the \llvmir
representation~\cite{grech15}.
The advantage is that this approach can be integrated more directly in the LLVM
toolchain; in principle it is applicable to any languages targeting LLVM. The
energy model used is exactly the same as the one applied
in~\cite{isa-vs-llvm-fopara} and described in \cref{ss:xcore-e-model}. 

\subsection{WCET-inspired energy consumption static analysis}
\label{sec:ecsa}

Since the underlying challenges of analysing the timing and energy
consumption behaviour of a program appear to be quite similar,
in~\cite{Georgiou2015arXiv}, 
we have applied well known WCET analysis techniques to retrieve
energy consumption estimations. One of the most popular WCET
techniques is implicit path enumeration (IPET)~\cite{Li1997}, which retrieves the worst case
control flow path of programs based on a cost model that assigns
a timing cost to each CFG basic block. We have replaced the
timing cost model by the ISA energy model given in~\Cref{eq:coremodel}. In the
absence of architectural performance enhancing features, such as caches, this
technique can provide safe upper bounds for timing. Through our
experimental evaluation we have demonstrated that this is not the case for
energy, as energy consumption, in contrast to time, is data
sensitive (see~\cref{conclusion}). 

In order to explore the value and limits of applying IPET for energy
consumption estimations, we have also extended the analysis to the \llvmir
level, using the \llvmir energy characterization given by the mapping technique
referred to in~\cref{subsec:mapping}. Furthermore, we have extended the energy
consumption analysis to multi-threaded embedded programs from two commonly used
concurrency patterns: task farms and pipelines. The experimental evaluation on
a set of mainly industrial programs demonstrates that, although the energy
bounds retrieved cannot be considered safe, they can still provide valuable
information for energy aware development, delivering energy transparency to
software developers in the absence of widely accessible software energy
monitoring.

\subsection{Probabilistic resource analysis}
\label{sec:probabilistic-analysis}
 
Bounds on energy consumption are useful, but
information about the distribution of consumption within those bounds
is even more so.  For example, it may be that most execution cases of
a program result in consumption close to the lower bound, while the
upper bound is reached only in a few outlying cases, or vice versa.
From the distribution, estimates of average energy consumption can be
derived.
One approach to obtaining this kind of information
is to perform probabilistic static analysis of a program with respect
to its energy consumption.  This is a special case of probabilistic
output analysis, whose aim is to derive a probability distribution of
possible output values for a program from a probability distribution
of its input. The output in this case is energy
consumption~\cite{DBLP:journals/corr/RosendahlK15,prob-res-fopara}.

\section{Discussion}\label{conclusion}
Energy-aware software development needs energy transparency;
designers and programmers have to understand  energy consumption
at an early stage in the development lifecycle in order to explore the
design space taking energy consumption into consideration.  Many decisions
taken early in the process, such as the hardware platform, the degree of parallelism,
fundamental algorithms and data structures, can determine the overall
energy efficiency of the final application. The energy-aware software development
lifecycle includes activities such as:
\begin{itemize}
\item
Providing an energy specification or energy budget.
\item
Making initial rough estimates of energy consumption based on high-level
models of the application, allowing exploration of design space with respect to
energy consumption.
\item
Choosing and configuring a hardware platform which suits the application, for
example reducing the energy cost of frequent communications or memory
accesses.
\item
Developing code with constant reference to the energy consumption of program
parts, allowing energy ``bugs" to be identified early.
\item
Performing energy optimisation of critical code sections using more precise energy models 
and taking into account the compiler generated machine instructions.
\item
For energy critical applications, providing guarantees in the form of tight
upper and lower bounds on energy consumption.
\end{itemize}

It is important to note that the development platform is seldom the
same as the final deployment
platform, emphasising the importance of energy modelling of the final target
hardware.  
The alternative to energy 
transparency is to wait until the application is run on the final intended platform and
then measure its energy usage. At this stage it is likely to be too late to
do much about excessive energy consumption.

\subsection{Software energy modelling challenges}

Verification of worst-case energy
consumption requires the development of a worst-case energy model.
This is difficult, 
since the energy cost of
executing an instruction 
depends on the operands. 
To obtain the worst-case consumption for an instruction we must
therefore measure its execution with the operands that induce
it. The energy models built in ENTRA are
based on averages obtained from measuring the energy consumed when
random, valid data is being processed. 
We demonstrated that the variation due to data can
range from 5 to 25\%~\cite{Georgiou2015arXiv}.
In~\cite{pallister2015data} we examine the impact of operand values on
instruction level energy consumption and propose a probabilistic
approach to developing worst-case energy models suitable for safe worst-case energy consumption analysis.

Data-sensitive energy modelling is a serious challenge. 
Determining the maximum amount of circuit switching between instruction data
tends to take an exponential amount of time to evaluate, making it difficult to
determine or guarantee the worst case data for an instruction
sequence~\cite{2016arXiv160302580M}. Even if a model is built where the energy
consumed by an instruction is given as a function of its operands, there would
still be further challenges in encoding these functions in a suitable way to be
exploited effectively by static analysis algorithms.

In~\cite{basic-block-energy-hip3es2016} we have explored a technique
that models both, upper and lower bounds on the energy of
``branchless'' blocks of instructions, in order to take into account
the inter-instruction switching costs within a block. It uses an
Evolutionary Algorithm for a faster exploration of the search space,
which is also reduced by the fact that the algorithm does not have to
deal with the control-flow of the program. Then, such block-level
model is fed into a static analysis, which takes into account the
program control-flow, and infers energy information for the whole
program and its procedures.

\subsection{Software energy analysis challenges}

Static analysis always involves a trade-off of precision against
complexity of the analysis.
Obtaining tight
bounds of energy usage depends on several factors, including accurate
propagation of
data size measures and extraction and solution of the relations
expressing energy consumption in terms of data sizes.  Both of these
problems are solvable for a large class of useful programs, but if
program structure departs from standard patterns, precision may be
rapidly lost.
For instance, our realisation of the general framework described in
Section~\ref{sec:generality} using the~\ciaopp resource analyser
described in Section~\ref{sec:horn-clause-analysis} (which uses the
\hcir) can deal with recursive programs, including multiple and mutual
recursion (e.g., divide-and-conquer programs), iterative programs with
nested loops, numeric/arithmetic programs, or programs operation on
complex data structures such as nested lists and arrays. The analysis
produces parametrised energy bounds (which depend on input data sizes)
expressed by a large class of functions (e.g., polynomial, factorial,
exponential, logarithmic, and summations), in contrast to other
approaches that are limited to polynomial functions, or to
non-parametric (i.e., absolute)
bounds.  

The experiments reported in~\cite{isa-vs-llvm-fopara} with this
realisation of the analysis framework perform the analysis at the \isa
and \llvmir \levels, and compare the energy values obtained by
evaluating the inferred energy functions for different input data
sizes, with actual hardware measurements (on the \xcore platform).
The results show that our \llvmir \level analysis is reasonably
accurate (less than $6.4\%$ average deviation vs.\ hardware
measurements) and more powerful than the analysis at the \isa \level,
in the sense that it can deal with a larger class of programs (e.g.,
programs involving structured types). The average deviation for the
smaller set of benchmarks for which the \isa-\level analysis produced
non-trivial results was 3.9\%.
Although we have tested our prototype tools with relatively small
programs, they exhibit features that are also present in bigger real
programs, and could be analysed at a bigger scale too, since we have
designed our analysis tools to enable scalability. Thus, we interpret
our experimental results as very promising, and encourage us to
continue our research following an incremental approach. Making our
prototype tools more robust, powerful and scalable, as well as
evaluating them with bigger real programs, is an implementation (and
research) challenge in itself.

As already said, our approach to developing energy analysers is
architecture neutral, a claim that is supported by the experimental
results in~\cite{grech15}, performed with the direct energy analysis
of \llvmir described in Section~\ref{sec:llvm-static-direct}, for both
the ARM Cortex-M and XMOS \xcore platforms. The benchmarks also
contain nested loops (some of them with complex control flow
predicates) and perform bitwise operations, as well as operations on
arrays and matrices. Overall, the final deviation vs.\ hardware
measurements is typically less than 10\% and 20\% on the XMOS and ARM
platforms respectively, showing that the general trend of the static
analysis results can be relied upon to give an estimate of the energy
consumption. Instantiating our general approach to more platforms
(which include the hardware and operating system) and assessing it in
different application domains is another challenge that we intend to
address in future work.

Static analysis of multi-threaded code is difficult
since precision is easily lost due to thread
interleaving. 
Accurate analysis of the timing and
synchronisation behaviour of threads is a pre-requisite for energy
analysis using the model given in \cref{eq:coremodel}, in which the
energy of an instruction depends on how many threads are active
simultaneously.

\subsection{Extending results to less predictable architectures}

The ENTRA approach is generic and architecture-neutral; that is, it
consists of a framework parametrised by an energy model and generic
static analysis tools.  Front-ends that translate code into a common,
analysable form such as \hcir or \llvmir can in principle be
constructed for any programming language. However, much
experimentation and investigation remains to be done to apply the
approach effectively to architectures that contain sources of
unpredictability
such
as caches, complex pipelines and interrupts that are not present in the
\xcore architecture.
A likely path of research is to follow the approach in WCET analysis
of such unpredictable architectures, employing supporting analyses to
permit more accurate energy analyses. An example is approximation of
cache contents at specific program points, leading to more accurate
models when it can be guaranteed that a memory access will definitely
hit or definitely miss the cache.

\subsection{Energy optimisations enabled by energy transparency.}

Different types of optimisations at different levels of the software
stack can be performed by taking advantage of the information provided
by the multi-level and multi-language ENTRA tools.
Both static and
dynamic energy optimisations are enabled by energy transparency,
which have been investigated in ENTRA.
For dynamic optimisations, a framework for energy-aware stochastic
scheduling based on evolutionary algorithms (EAs) has been developed,
for the cases where tasks are independent~\cite{scheduling-neucom14}
and dependent~\cite{bankovic-copula:2015}. In the latter case,
dependence has been modelled by using copula theory~\cite{Nelsen}, in
particular Archimedean copulas~\cite{McNeil}.
EAs have also been used to improve energy-aware allocation and
scheduling for DVFS-enabled multicore environments. For example, the
algorithms described in~\cite{bankovic-ga:2015,bankovic-aiai:2015} are
able to deal with task migration and preemption; and the ones
in~\cite{bankovic-loop_perf:2015} allow decreasing program accuracy
(by performing loop perforation) in order to save energy.

Other optimisations include the use of energy analysis to choose
software parameters in order to transform programs to ensure that an
energy target is met while minimizing the loss in quality of service.
Energy performance has also been improved by optimisation techniques
for task placement on an \xcore network after identifying
communication patterns among tasks. Using Swallow, an experimental
open platform of many xCOREs~\cite{KerrisonSwallow15} as a source of
model data, it has been demonstrated that incorporating network-level
energy consumption and timing into the energy modelling process can
aid in identifying the impact of sub-optimal task placement in
communicating multi-core
applications~\cite{DBLP:journals/corr/KerrisonE15a}.

Some energy optimisations have already been incorporated into the
recently released XC compiler by XMOS Ltd. These optimisations in the
object code are obtained by aggressively applying global dataflow
analyses inspired by ENTRA project research. Results on case studies,
showing a power reduction of approximately 25\% by using the global optimiser,
can be found in an ENTRA project report~\cite{entra-d6.2}.

An experiment on energy optimisation of Android code was carried out using the
source code energy modelling mentioned in Section \ref{subsec:mapping}.
The energy consumption of battery-driven mobile devices such as smartphones is of 
increasing concern to developers of software for
such devices.  The study concerned  the optimisation of code for interactive games.
The energy model combined with execution profiling, 
enabled the developer to discover that 10 code blocks (out of several hundred blocks in
the considered code) consumed over 50\% of the overall energy. Aggressive source code
optimisation and refactoring of these blocks enabled energy savings of 6\% to 50\% in various
use-case scenarios \cite{LiGallagher-TSE-2016}.

\section{Conclusions}

The goal of the ENTRA project was to provide techniques and tools supporting energy transparency
at the software level. The results obtained include energy models for
the \xcore processor at different levels of abstraction, from ISA level
to \llvmir, as well as preliminary energy models for less predictable architectures.
A model incorporating multi-core execution on the \xcore has
also been developed. These models are incorporated in static analyses
at the corresponding level of code. Experiments have compared
predictions of energy consumption for single-threaded programs with
actual measured consumption, and encouraging results with only a few
percentage of error were obtained.  In addition, the trade-offs between
accuracy and analysability at different levels were explored, leading
to a preliminary conclusion that analysis at LLVM level provides a
good compromise. The project identified challenging problems for
future research,  extending the analyses to multi-threaded code, and building data-sensitive
energy models.

\section*{Acknowledgements}
This research 
has received funding from the European Union 7th Framework Programme
under grant agreement no.\  318337, ENTRA - Whole Systems Energy
Transparency, grant agreement no.\  611004, the ICT-Energy Coordination
and Support Action, Spanish MINECO TIN'12-39391 \emph{StrongSoft}
project, and the Madrid M141047003 \emph{N-GREENS} program.

% \section*{References}

% \bibliography{./rc,./entra}

\begin{thebibliography}{10}
\expandafter\ifx\csname url\endcsname\relax
  \def\url#1{\texttt{#1}}\fi
\expandafter\ifx\csname urlprefix\endcsname\relax\def\urlprefix{URL }\fi
\expandafter\ifx\csname href\endcsname\relax
  \def\href#1#2{#2} \def\path#1{#1}\fi

\bibitem{Edwards:lssmlps}
C.~Edwards., Lack of software support marks the low power scorecard at {DAC},
  in: Electronics Weekly., 2011.

\bibitem{Roy_Johnson_1997}
K.~Roy, M.~C. Johnson,
  \href{http://citeseerx.ist.psu.edu/viewdoc/summary?doi=10.1.1.44.4398}{{S}oftware
  {D}esign for {L}ow {P}ower}, in: W.~Nebel, J.~P. Mermet (Eds.), Low Power
  Design in Deep Submicron Electronics, Vol. 337, Kluwer Academic, 1997, pp.
  433--460.
\newline\urlprefix\url{http://citeseerx.ist.psu.edu/viewdoc/summary?doi=10.1.1.44.4398}

\bibitem{wattch:2000}
D.~Brooks, V.~Tiwari, M.~Martonosi,
  \href{http://doi.acm.org/10.1145/339647.339657}{Wattch: a framework for
  architectural-level power analysis and optimizations}, in: Proceedings of the
  27th annual international symposium on Computer architecture, ISCA '00, ACM,
  New York, NY, USA, 2000, pp. 83--94.
\newblock \href {http://dx.doi.org/http://doi.acm.org/10.1145/339647.339657}
  {\path{doi:http://doi.acm.org/10.1145/339647.339657}}.
\newline\urlprefix\url{http://doi.acm.org/10.1145/339647.339657}

\bibitem{TiwariWolfeInstructionLevelPowerAnalysi:1996}
V.~Tiwari, S.~Malik, A.~Wolfe, M.~Tien-Chien~Lee,
  \href{http://dx.doi.org/10.1007/BF01130407}{Instruction level power analysis
  and optimization of software}, The Journal of VLSI Signal Processing 13
  (1996) 223--238, 10.1007/BF01130407.
\newline\urlprefix\url{http://dx.doi.org/10.1007/BF01130407}

\bibitem{Steinke2001}
S.~Steinke, M.~Knauer, L.~Wehmeyer, P.~Marwedel, {An Accurate and Fine Grain
  Instruction-level Energy Model Supporting Software Optimizations}, in:
  Proceedings of PATMOS, 2001.

\bibitem{DBLP:journals/tecs/KerrisonE15}
S.~Kerrison, K.~Eder, \href{http://doi.acm.org/10.1145/2700104}{Energy modeling
  of software for a hardware multithreaded embedded microprocessor}, {ACM}
  Trans. Embedded Comput. Syst. 14~(3) (2015) 56.
\newblock \href {http://dx.doi.org/10.1145/2700104}
  {\path{doi:10.1145/2700104}}.
\newline\urlprefix\url{http://doi.acm.org/10.1145/2700104}

\bibitem{Brandolese2011}
C.~Brandolese, S.~Corbetta, W.~Fornaciari, Software energy estimation based on
  statistical characterization of intermediate compilation code, in: Low Power
  Electronics and Design (ISLPED) 2011 International Symposium on, 2011, pp.
  333--338.
\newblock \href {http://dx.doi.org/10.1109/ISLPED.2011.5993659}
  {\path{doi:10.1109/ISLPED.2011.5993659}}.

\bibitem{Georgiou2015arXiv}
K.~{Georgiou}, S.~{Kerrison}, K.~{Eder},
  \href{http://arxiv.org/abs/1510.07095}{{On the Value and Limits of
  Multi-level Energy Consumption Static Analysis for Deeply Embedded Single and
  Multi-threaded Programs}}, Tech. rep. (Oct. 2015).
\newblock \href {http://arxiv.org/abs/1510.07095} {\path{arXiv:1510.07095}}.
\newline\urlprefix\url{http://arxiv.org/abs/1510.07095}

\bibitem{DBLP:journals/cacm/Wegbreit75}
B.~Wegbreit, Mechanical program analysis, Commun. ACM 18~(9) (1975) 528--539.

\bibitem{Rosendahl89}
M.~Rosendahl, {A}utomatic {C}omplexity {A}nalysis, in: 4th ACM {C}onference on
  {F}unctional {P}rogramming {L}anguages and {C}omputer {A}rchitecture
  (FPCA'89), ACM Press, 1989, pp. 144--156.

\bibitem{granularity}
S.~K. Debray, N.-W. Lin, M.~Hermenegildo, {T}ask {G}ranularity {A}nalysis in
  {L}ogic {P}rograms, in: Proc. of the 1990 {ACM} Conf. on Programming Language
  Design and Implementation, {ACM} Press, 1990, pp. 174--188.

\bibitem{low-bounds-ilps97}
S.~K. Debray, P.~L\'{o}pez-Garc\'{i}a, M.~Hermenegildo, N.-W. Lin, {L}ower
  {B}ound {C}ost {E}stimation for {L}ogic {P}rograms, in: 1997 International
  Logic Programming Symposium, MIT Press, Cambridge, MA, 1997, pp. 291--305.

\bibitem{resource-iclp07}
J.~Navas, E.~Mera, P.~L\'{o}pez-Garc\'{i}a, M.~Hermenegildo, {U}ser-{D}efinable
  {R}esource {B}ounds {A}nalysis for {L}ogic {P}rograms, in: International
  Conference on Logic Programming (ICLP'07), Lecture Notes in Computer Science,
  Springer, 2007, pp. 348--363.

\bibitem{NMHLFM08}
J.~Navas, M.~M\'{e}ndez-Lojo, M.~Hermenegildo, {S}afe {U}pper-bounds
  {I}nference of {E}nergy {C}onsumption for {J}ava {B}ytecode {A}pplications,
  in: The Sixth NASA Langley Formal Methods Workshop (LFM 08), 2008, pp.
  29--32, extended Abstract.

\bibitem{plai-resources-iclp14}
A.~Serrano, P.~Lopez-Garcia, M.~Hermenegildo,
  \href{http://arxiv.org/abs/1405.4256}{{R}esource {U}sage {A}nalysis of
  {L}ogic {P}rograms via {A}bstract {I}nterpretation {U}sing {S}ized {T}ypes},
  Theory and Practice of Logic Programming, 30th Int'l. Conference on Logic
  Programming (ICLP'14) Special Issue 14~(4-5) (2014) 739--754.
\newblock \href {http://dx.doi.org/10.1017/S147106841400057X}
  {\path{doi:10.1017/S147106841400057X}}.
\newline\urlprefix\url{http://arxiv.org/abs/1405.4256}

\bibitem{AlbertAGP11a}
E.~Albert, P.~Arenas, S.~Genaim, G.~Puebla, {C}losed-{F}orm {U}pper {B}ounds in
  {S}tatic {C}ost {A}nalysis, Journal of Automated Reasoning 46~(2) (2011)
  161--203.

\bibitem{XS1-Architecture}
D.~May, The {XMOS} {XS1} architecture. available online:
  http://www.xmos.com/published/xmos-xs1-architecture (2013).

\bibitem{Tiwari1994a}
V.~Tiwari, S.~Malik, A.~Wolfe, {Compilation Techniques for Low Energy: An
  Overview}, in: Low Power Electronics, 1994. Digest of Technical Papers., IEEE
  Symposium, IEEE, 1994, pp. 38--39.

\bibitem{pallister2015data}
J.~Pallister, S.~Kerrison, J.~Morse, K.~Eder,
  \href{http://arxiv.org/abs/1505.03374}{Data dependent energy modeling for
  worst case energy consumption analysis}, Tech. rep. (May 2015).
\newblock \href {http://arxiv.org/abs/1505.03374} {\path{arXiv:1505.03374}}.
\newline\urlprefix\url{http://arxiv.org/abs/1505.03374}

\bibitem{2015arXiv150902830K}
S.~{Kerrison}, K.~{Eder}, \href{http://arxiv.org/abs/1509.02830}{{Modeling and
  visualizing networked multi-core embedded software energy consumption}},
  Tech. rep. (Sep. 2015).
\newblock \href {http://arxiv.org/abs/1509.02830} {\path{arXiv:1509.02830}}.
\newline\urlprefix\url{http://arxiv.org/abs/1509.02830}

\bibitem{influence-wilhelm-2003}
R.~Heckmann, M.~Langenbach, S.~Thesing, R.~Wilhelm, The influence of processor
  architecture on the design and the results of {WCET} tools, Proceedings of
  the IEEE 91~(7) (2003) 1038--1054.
\newblock \href {http://dx.doi.org/10.1109/JPROC.2003.814618}
  {\path{doi:10.1109/JPROC.2003.814618}}.

\bibitem{isa-energy-lopstr13-final}
U.~Liqat, S.~Kerrison, A.~Serrano, K.~Georgiou, P.~Lopez-Garcia, N.~Grech,
  M.~Hermenegildo, K.~Eder,
  \href{http://dx.doi.org/10.1007/978-3-319-14125-1_5}{{E}nergy {C}onsumption
  {A}nalysis of {P}rograms based on {XMOS} {ISA}-level {M}odels}, in: G.~Gupta,
  R.~Peña (Eds.), Logic-Based Program Synthesis and Transformation, 23rd
  International Symposium, {LOPSTR} 2013, Revised Selected Papers, Vol. 8901 of
  Lecture Notes in Computer Science, Springer, 2014, pp. 72--90.
\newblock \href {http://dx.doi.org/10.1007/978-3-319-14125-1_5}
  {\path{doi:10.1007/978-3-319-14125-1_5}}.
\newline\urlprefix\url{http://dx.doi.org/10.1007/978-3-319-14125-1_5}

\bibitem{LattnerLLVM2004}
C.~Lattner, V.~Adve,
  \href{http://dl.acm.org/citation.cfm?id=977395.977673}{{LLVM}: A compilation
  framework for lifelong program analysis and transformation}, in: Proc. of the
  2004 International Symposium on Code Generation and Optimization (CGO), IEEE
  Computer Society, 2004, pp. 75--88.
\newline\urlprefix\url{http://dl.acm.org/citation.cfm?id=977395.977673}

\bibitem{2015arXiv151004165L}
X.~Li, J.~P. Gallagher, \href{http://arxiv.org/pdf/1510.04165v1.pdf}{A
  top-to-bottom view: Energy analysis for mobile application source code},
  Tech. rep., submitted to a conference (October 2015).
\newblock \href {http://arxiv.org/abs/1510.04165} {\path{arXiv:1510.04165}}.
\newline\urlprefix\url{http://arxiv.org/pdf/1510.04165v1.pdf}

\bibitem{isa-vs-llvm-fopara}
U.~Liqat, K.~Georgiou, S.~Kerrison, P.~Lopez-Garcia, M.~V. Hermenegildo, J.~P.
  Gallagher, K.~Eder, \href{http://arxiv.org/abs/1511.01413}{{I}nferring
  {P}arametric {E}nergy {C}onsumption {F}unctions at {D}ifferent {S}oftware
  {L}evels: {ISA} vs. {LLVM IR}}, in: M.~V. Eekelen, U.~D. Lago (Eds.),
  Foundational and Practical Aspects of Resource Analysis. Fourth International
  Workshop FOPARA 2015, Revised Selected Papers, Lecture Notes in Computer
  Science, Springer, 2016, in press.
\newblock \href {http://arxiv.org/abs/1511.01413} {\path{arXiv:1511.01413}}.
\newline\urlprefix\url{http://arxiv.org/abs/1511.01413}

\bibitem{staticprofiling-flops}
R.~Haemmerl{\'e}, P.~Lopez-Garcia, U.~Liqat, M.~Klemen, J.~P. Gallagher, M.~V.
  Hermenegildo, \href{http://dx.doi.org/10.1007/978-3-319-29604-3_11}{{A}
  {T}ransformational {A}pproach to {P}arametric {A}ccumulated-cost {S}tatic
  {P}rofiling}, in: O.~Kiselyov, A.~King (Eds.), 13th International Symposium
  on Functional and Logic Programming (FLOPS 2016), Vol. 9613 of LNCS,
  Springer, 2016, pp. 163--180.
\newblock \href {http://dx.doi.org/10.1007/978-3-319-29604-3_11}
  {\path{doi:10.1007/978-3-319-29604-3_11}}.
\newline\urlprefix\url{http://dx.doi.org/10.1007/978-3-319-29604-3_11}

\bibitem{HenriksenG06}
K.~S. Henriksen, J.~P. Gallagher, Abstract interpretation of {PIC} programs
  through logic programming, in: Sixth IEEE International Workshop on Source
  Code Analysis and Manipulation (SCAM 2006), IEEE Computer Society, 2006, pp.
  184--196.

\bibitem{decomp-oo-prolog-lopstr07}
M.~M\'{e}ndez-Lojo, J.~Navas, M.~Hermenegildo, {A} {F}lexible ({C}){LP}-{B}ased
  {A}pproach to the {A}nalysis of {O}bject-{O}riented {P}rograms, in: 17th
  International Symposium on Logic-based Program Synthesis and Transformation
  (LOPSTR 2007), no. 4915 in LNCS, Springer-Verlag, 2007, pp. 154--168.

\bibitem{DBLP:conf/tacas/GrebenshchikovGLPR12}
S.~Grebenshchikov, A.~Gupta, N.~P. Lopes, C.~Popeea, A.~Rybalchenko, {HSF(C)}:
  {A} {S}oftware {V}erifier {B}ased on {H}orn {C}lauses - ({C}ompetition
  {C}ontribution), in: C.~Flanagan, B.~K{\"o}nig (Eds.), TACAS, Vol. 7214 of
  LNCS, Springer, 2012, pp. 549--551.

\bibitem{DBLP:conf/fm/HojjatKGIKR12}
H.~Hojjat, F.~Konecn{\'{y}}, F.~Garnier, R.~Iosif, V.~Kuncak, P.~R{\"{u}}mmer,
  \href{http://dx.doi.org/10.1007/978-3-642-32759-9_21}{{A} {V}erification
  {T}oolkit for {N}umerical {T}ransition {S}ystems - {T}ool {P}aper}, in:
  D.~Giannakopoulou, D.~M{\'{e}}ry (Eds.), {FM} 2012: Formal Methods - 18th
  International Symposium, Paris, France, August 27-31, 2012. Proceedings, Vol.
  7436 of Lecture Notes in Computer Science, Springer, 2012, pp. 247--251.
\newblock \href {http://dx.doi.org/10.1007/978-3-642-32759-9_21}
  {\path{doi:10.1007/978-3-642-32759-9_21}}.
\newline\urlprefix\url{http://dx.doi.org/10.1007/978-3-642-32759-9_21}

\bibitem{z3}
L.~M. de~Moura, N.~Bj{\o}rner, {Z3}: {A}n {E}fficient {SMT} {S}olver, in: C.~R.
  Ramakrishnan, J.~Rehof (Eds.), Tools and Algorithms for the Construction and
  Analysis of Systems, 14th International Conference, TACAS 2008, Vol. 4963 of
  Lecture Notes in Computer Science, Springer, 2008, pp. 337--340.

\bibitem{hcvs14}
N.~Bj{\o}rner, F.~Fioravanti, A.~Rybalchenko, V.~Senni (Eds.),
  \href{http://dx.doi.org/10.4204/EPTCS.169}{{P}roceedings of {F}irst
  {W}orkshop on {H}orn {C}lauses for {V}erification and {S}ynthesis}, Vol. 169
  of {EPTCS}, 2014.
\newblock \href {http://dx.doi.org/10.4204/EPTCS.169}
  {\path{doi:10.4204/EPTCS.169}}.
\newline\urlprefix\url{http://dx.doi.org/10.4204/EPTCS.169}

\bibitem{hermenegildo11:ciao-design-tplp}
M.~V. Hermenegildo, F.~Bueno, M.~Carro, P.~L\'{o}pez, E.~Mera, J.~Morales,
  G.~Puebla, \href{http://arxiv.org/abs/1102.5497}{{A}n {O}verview of {C}iao
  and its {D}esign {P}hilosophy}, Theory and Practice of Logic Programming
  12~(1--2) (2012) 219--252.
\newblock \href {http://dx.doi.org/doi:10.1017/S1471068411000457}
  {\path{doi:doi:10.1017/S1471068411000457}}.
\newline\urlprefix\url{http://arxiv.org/abs/1102.5497}

\bibitem{entra-d2.1}
K.~Eder, N.~Grech (Eds.), Common Assertion Language, ENTRA Project:
  Whole-Systems Energy Transparency (FET project 318337), 2013, deliverable
  2.1, http://entraproject.eu.

\bibitem{ai-jlp}
K.~Muthukumar, M.~Hermenegildo, {C}ompile-time {D}erivation of {V}ariable
  {D}ependency {U}sing {A}bstract {I}nterpretation, Journal of Logic
  Programming 13~(2/3) (1992) 315--347.

\bibitem{grech15}
N.~Grech, K.~Georgiou, J.~Pallister, S.~Kerrison, J.~Morse, K.~Eder,
  \href{http://dl.acm.org/citation.cfm?doid=2764967.2764974}{Static analysis of
  energy consumption for {LLVM IR} programs}, in: Proceedings of the 18th
  International Workshop on Software and Compilers for Embedded Systems, SCOPES
  2015, ACM, New York, NY, USA, 2015, pp. 12--21.
\newblock \href {http://dx.doi.org/10.1145/2764967.2764974}
  {\path{doi:10.1145/2764967.2764974}}.
\newline\urlprefix\url{http://dl.acm.org/citation.cfm?doid=2764967.2764974}

\bibitem{Li1997}
Y.-S. Li, S.~Malik, Performance analysis of embedded software using implicit
  path enumeration, Computer-Aided Design of Integrated Circuits and Systems,
  IEEE Transactions on 16~(12) (1997) 1477--1487.
\newblock \href {http://dx.doi.org/10.1109/43.664229}
  {\path{doi:10.1109/43.664229}}.

\bibitem{DBLP:journals/corr/RosendahlK15}
M.~Rosendahl, M.~H. Kirkeby,
  \href{http://dx.doi.org/10.4204/EPTCS.194.8}{Probabilistic output analysis by
  program manipulation}, in: N.~Bertrand, M.~Tribastone (Eds.), Proceedings
  Thirteenth Workshop on Quantitative Aspects of Programming Languages and
  Systems, {QAPL} 2015, London, UK, 11th-12th April 2015., Vol. 194 of {EPTCS},
  2015, pp. 110--124, qAPL, London.
\newblock \href {http://dx.doi.org/10.4204/EPTCS.194.8}
  {\path{doi:10.4204/EPTCS.194.8}}.
\newline\urlprefix\url{http://dx.doi.org/10.4204/EPTCS.194.8}

\bibitem{prob-res-fopara}
M.~Kirkeby, M.~Rosendahl,
  \href{http://entraproject.eu/wp-content/uploads/2016/02/prob-res-fopara.pdf}{{P}robabilistic
  {R}esource {A}nalysis by {P}rogram {T}ransformation}, in: Proc. of the
  Foundational and Practical Aspects of Resource Analysis, LNCS, Springer,
  2015, in press.
\newline\urlprefix\url{http://entraproject.eu/wp-content/uploads/2016/02/prob-res-fopara.pdf}

\bibitem{2016arXiv160302580M}
J.~{Morse}, S.~{Kerrison}, K.~{Eder},
  \href{http://arxiv.org/abs/1603.02580}{{On the infeasibility of analysing
  worst-case dynamic energy}}, Tech. rep. (Mar. 2016).
\newblock \href {http://arxiv.org/abs/1603.02580} {\path{arXiv:1603.02580}}.
\newline\urlprefix\url{http://arxiv.org/abs/1603.02580}

\bibitem{basic-block-energy-hip3es2016}
U.~Liqat, Z.~Bankovi\'{c}, P.~Lopez-Garcia, M.~V. Hermenegildo,
  \href{http://arxiv.org/abs/1601.02800}{{I}nferring {E}nergy {B}ounds
  {S}tatically by {E}volutionary {A}nalysis of {B}asic {B}locks}, in: Workshop
  on High Performance Energy Efficient Embedded Systems (HIP3ES 2016), 2016,
  arXiv:1601.02800.
\newline\urlprefix\url{http://arxiv.org/abs/1601.02800}

\bibitem{scheduling-neucom14}
Z.~Bankovi\'{c}, P.~Lopez-Garcia,
  \href{http://www.sciencedirect.com/science/article/pii/S0925231214012211}{{S}tochastic
  vs. {D}eterministic {E}volutionary {A}lgorithm-based {A}llocation and
  {S}cheduling for {XMOS} {C}hips}, Neurocomputing 150 (2015) 82--89.
\newblock \href
  {http://dx.doi.org/http://dx.doi.org/10.1016/j.neucom.2014.06.077}
  {\path{doi:http://dx.doi.org/10.1016/j.neucom.2014.06.077}}.
\newline\urlprefix\url{http://www.sciencedirect.com/science/article/pii/S0925231214012211}

\bibitem{bankovic-copula:2015}
Z.~Bankovi\'{c}, P.~L\'{o}pez-Garc\'{i}a,
  \href{http://dx.doi.org/10.1007/978-3-319-19719-7_14}{{I}mproved
  {E}nergy-aware {S}tochastic {S}cheduling based on {E}volutionary {A}lgorithms
  via {C}opula-based {M}odeling of {T}ask {D}ependences}, in: A.~Herrero,
  J.~Sedano, B.~Baruque, H.~Quinti\'{a}n, E.~Corchado (Eds.), International
  Conference on Soft Computing Models in Industrial and Environmental
  Applications (SOCO 2015), Vol. 368 of Advances in Intelligent Systems and
  Computing, Springer International Publishing, 2015, pp. 153--163.
\newblock \href {http://dx.doi.org/10.1007/978-3-319-19719-7_14}
  {\path{doi:10.1007/978-3-319-19719-7_14}}.
\newline\urlprefix\url{http://dx.doi.org/10.1007/978-3-319-19719-7_14}

\bibitem{Nelsen}
R.~B. Nelsen, Properties and applications of copulas: A brief survey, in: First
  Brazilian Conference on Statistical Modelling in Insurance and Finance, 2003,
  pp. 10--28.

\bibitem{McNeil}
A.~J. McNeil, et~al., Multivariate archimedean copulas, $d$-monotone functions
  and $l_1$-norm symmetric distributions (2009).

\bibitem{bankovic-ga:2015}
Z.~Bankovi\'{c}, P.~L\'{o}pez-Garc\'{i}a,
  \href{http://doi.acm.org/10.1145/2739482.2764645}{{E}nergy {E}fficient
  {A}llocation and {S}cheduling for {DVFS}-enabled {M}ulticore {E}nvironments
  using a {M}ultiobjective {E}volutionary {A}lgorithm}, in: Genetic and
  Evolutionary Computation Conference, {GECCO} 2015, Companion Material
  Proceedings, ACM, 2015, pp. 1353--1354.
\newblock \href {http://dx.doi.org/10.1145/2739482.2764645}
  {\path{doi:10.1145/2739482.2764645}}.
\newline\urlprefix\url{http://doi.acm.org/10.1145/2739482.2764645}

\bibitem{bankovic-aiai:2015}
Z.~Bankovi\'{c}, U.~Liqat, P.~L\'{o}pez-Garc\'{i}a,
  \href{http://dx.doi.org/10.1007/978-3-319-23868-5_35}{{A} {P}ractical
  {A}pproach for {E}nergy {E}fficient {S}cheduling in {M}ulticore
  {E}nvironments by combining {E}volutionary and {YDS} {A}lgorithms with
  {F}aster {E}nergy {E}stimation}, in: The 11th International Conference on
  Artificial Intelligence Applications and Innovations (AIAI'15), Vol. 458 of
  IFIP Advances in Information and Communication Technology, Springer
  International Publishing, 2015, pp. 478--493.
\newblock \href {http://dx.doi.org/10.1007/978-3-319-23868-5_35}
  {\path{doi:10.1007/978-3-319-23868-5_35}}.
\newline\urlprefix\url{http://dx.doi.org/10.1007/978-3-319-23868-5_35}

\bibitem{bankovic-loop_perf:2015}
Z.~Bankovi\'{c}, U.~Liqat, P.~L\'{o}pez-Garc\'{i}a,
  \href{http://link.springer.com/chapter/10.1007/978-3-319-19644-2_57}{{T}rading-off
  {A}ccuracy vs. {E}nergy in {M}ulticore {P}rocessors via {E}volutionary
  {A}lgorithms {C}ombining {L}oop {P}erforation and {S}tatic {A}nalysis-based
  {S}cheduling}, in: E.~Onieva, I.~Santos, E.~Osaba, H.~Quinti\'{a}n,
  E.~Corchado (Eds.), Hybrid Artificial Intelligent Systems (HAIS 2015), Vol.
  9121 of Lecture Notes in Computer Science, Springer International Publishing,
  2015, pp. 690--701.
\newblock \href {http://dx.doi.org/10.1007/978-3-319-19644-2_57}
  {\path{doi:10.1007/978-3-319-19644-2_57}}.
\newline\urlprefix\url{http://link.springer.com/chapter/10.1007/978-3-319-19644-2_57}

\bibitem{KerrisonSwallow15}
S.~J. Hollis, S.~Kerrison,
  \href{http://entraproject.eu/wp-content/uploads/2016/02/KerrisonSwallow15.pdf}{{Swallow:
  Building an Energy-Transparent Many-Core Embedded Real-Time System}}, in:
  2016 Design, Automation \& Test in Europe, IEEE, 2016.
\newline\urlprefix\url{http://entraproject.eu/wp-content/uploads/2016/02/KerrisonSwallow15.pdf}

\bibitem{DBLP:journals/corr/KerrisonE15a}
S.~Kerrison, K.~Eder, \href{http://arxiv.org/abs/1509.02830}{Modeling and
  visualizing networked multi-core embedded software energy consumption}, CoRR
  abs/1509.02830.
\newline\urlprefix\url{http://arxiv.org/abs/1509.02830}

\bibitem{entra-d6.2}
H.~Muller (Ed.),
  \href{http://entraproject.eu/wp-content/uploads/2016/03/deliv_6.2.pdf}{Evaluation
  Results}, ENTRA Project: Whole-Systems Energy Transparency (FET project
  318337), 2016, deliverable 6.2, http://entraproject.eu.
\newline\urlprefix\url{http://entraproject.eu/wp-content/uploads/2016/03/deliv_6.2.pdf}

\bibitem{LiGallagher-TSE-2016}
X.~Li, J.~P. Gallagher,
  \href{http://entraproject.eu/wp-content/uploads/2016/02/LiGallagher-TSE-2016.pdf}{An
  energy-aware programming approach for mobile application development guided
  by a fine-grained energy model}, Tech. rep., Roskilde University, submitted
  for publication (February 2016).
\newline\urlprefix\url{http://entraproject.eu/wp-content/uploads/2016/02/LiGallagher-TSE-2016.pdf}

\end{thebibliography}

\end{document}